# Classifying Rhoticity of /ɹ/ in Speech Sound Disorder using Age-and-Sex Normalized Formants

*Nina R Benway[1], Jonathan L Preston[1,2], Asif Salekin[3], Yi Xiao[3], Harshit Sharma[3], Tara McAllister[4]*

[1]Communication Sciences and Disorders, Syracuse University, New York, USA
[2]Haskins Laboratories, New Haven, Connecticut, USA
[3]Electrical Engineering and Computer Science, Syracuse University, New York, USA
[4]Communicative Sciences and Disorders, New York University, New York, USA
nrbenway@syr.edu

## Abstract

Mispronunciation detection tools could increase treatment access for speech sound disorders impacting, e.g., /ɹ/. We show age-and-sex normalized formant estimation outperforms cepstral representation for detection of fully rhotic vs. derhotic /ɹ/ in the PERCEPT-R Corpus. Gated recurrent neural networks trained on this feature set achieve a mean test participant-specific F1-score =.81 ($\sigma_x$=.10, med = .83, n = 48), with post hoc modeling showing no significant effect of child age or sex.

## 1. Introduction

There are many access barriers to sufficiently-intense, evidence-based speech-language therapy services worldwide [1], including for childhood speech sound disorders. Clinical speech technologies—what we define as computerized versions of *clinically-validated evidence-based practices*—can potentially ameliorate this intensity gap, particularly when coupled with validated mispronunciation detection algorithms that drive clinical feedback and practice difficulty. Systematic review [2-5], however, has identified several barriers to the uptake of effective clinical speech technologies, particularly for child use: a paucity of public speech corpora for system training, inadequate technical description of tools, low accuracy when rating sounds in error, and insufficient clinical validation.

Recently published open-access data, the PERCEPT Corpora, begins to offset the lack of child speech corpora in the context of speech sound disorder [6], including for /ɹ/ (i.e., rhotics) in American English (the most common such error). Given the breadth and depth of perceptually-labeled /ɹ/ in the PERCEPT-R Corpus, and the constrained target phone, we are able to approach this task with a classifier which can outperform other algorithms (i.e., goodness of pronunciation) under these assumptions [7, 8]. We presently offer two contributions. First, a preliminary investigation demonstrates that age-and-sex normalization of formants outperforms Mel frequency cepstral coefficient features (MFCCs) for classification of fully rhotic/ derhotic /ɹ/. Second, we advance the test-participant-specific F1-score for rhotic mispronunciation detection in speech sound disorder literature from $\bar{x}$= .64 ($\sigma_x$= .25) [9] to $\bar{x}$= .81 ($\sigma_x$= .10; med = .83, n = 48). Clinical contributions are described in [10].

## 2. Automated Detection of Rhoticity

Automatic mispronunciation detection that generates clinically-valid feedback must presumably attend to the same perceptual features that a clinician would when rating /ɹ/. The work of [11] has shown that neural populations in the human superior temporal gyrus are sensitive to formant-indexed phonetic features, contrasting with MFCC features most commonly used in existing clinical mispronunciation detection systems [5].

### 2.1. State of the Art in Rhoticity Classification in Speech Sound Disorders

The few attempts at /ɹ/ classification in speech sound disorders have focused on ultrasound image features. Ribeiro and colleagues [9] used these features to train a convolutional neural network to classify typical versus mispronounced (i.e., glided) rhotics in Ultrasuite [12]. Reanalysis of the 8 test speakers with speech sound disorder in Table 2 of [9] suggests a participant-specific F1-score $\bar{x}$= .64 ($\sigma_x$ = .25). Note that the authors urge caution in interpreting the results because of low /ɹ/ perceptual ground truth agreement. Separately, Li and colleagues [13] trained support vector machines using automatically-generated quantifications of ultrasound images to classify accurate and misarticulated /ɹ/ in /ɑɹ/ for individuals with speech sound disorders in a private dataset. The authors report classification percent accuracies exceeding 89% in lab testing, but precise classification performance and participant-specific F1-score are not reported and hinder further interpretation.

### 2.2. Formant Structure for /ɹ/

Although these state-of-the-art baselines use MFCC features, much of the existing literature regarding the acoustic quantification of /ɹ/ uses formants. There is variation in the vocal tract configurations that generate a fully-rhotic /ɹ/, but such configurations yield spectral envelopes acoustically marked by a relatively high second formant (F2 [14]) and a relatively low third formant (F3; [15]). This results in the average F3-F2 distance of fully rhotic /ɹ/ being much narrower than the average F3-F2 distance produced by a neutral vocal tract, all else being equal, as can be interpolated from the norms generated by Lee and colleagues [16] (e.g., females, 10 years of age: F3-F2$_{rhotic}$ = 531 Hz, F3-F2$_{neutral}$ = 1683 Hz). However, because formant frequencies are influenced by vocal tract shape and size, the F3-F2 distances that are associated with perceptual correctness in, for example, 10-year-old females, are approximately one standard deviation above those in 14 year old males [16]. Furthermore, age-and-sex normalized F3-F2 has been found to better model expert listener ratings of fully rhotic versus derhotic /ɹ/ than raw F3-F2 distance [17], perhaps due to the ability of the normalized metric to account for differences arising for individual variation in size and shape of the vocal tract cavity resonator. Herein, we test the hypothesis that age-and-sex normalized formants features increase /ɹ/ classification performance compared to utterance-standardized formants and MFCC features.

# 3. Methods

## 3.1. Corpus Data

Data come from the PERCEPT-R Corpus v2.2.2, which includes the published open access subset (2.2.2p) [6] as well as participants excluded from the open access subset after review of consent/assent documents. General characteristics of the dataset are detailed extensively in other publications and data repositories. Briefly, PERCEPT-R Corpus v2.2.2 consists largely of single-word citation speech audio recorded during 27 longitudinal clinical trials of children with speech sound disorder impacting /ɹ/ and cross-sectional studies of age-match peers with typical speech. There are a total of 179,076 labeled utterances in the full PERCEPT-R Corpus v2.2.2 representing 662 single-rhotic words and phrases. Each audio file is matched to a ground-truth label for listener judgment of rhoticity. Ground-truth labels were generated by averaging binary listener ratings (0 = derhotic, 1 = fully rhotic) from several listeners. The heuristic for discretizing the average to binary ground-truth labels in the present investigation was .66. There is an imbalance in the source data favoring derhotic tokens (122,270: 56,806), as is expected from the corpus sources (clinical trials).

## 3.2. Dataset Design for Clinical Replicability

Because the intended use case for this mispronunciation detection system is clinical intervention, in order to maximize external validity and replicability of model performance we hand crafted the validation and test datasets to reflect the subset of participants in the PERCET-R Corpus for whom automated independent practice would be clinically appropriate. First, all speakers *without* speech sound disorder (i.e., the typically-speaking subset of PERCEPT-R) and speakers with /ɹ/ accuracy > .8 were excluded from the validation and test subsets. Second, because we believe that independent practice is only ethical for individuals who can produce a fully-rhotic /ɹ/ some of the time (i.e., *stimulable*), only speakers with a fully rhotic/derhotic proportion > .33 were prioritized for the validation and test subsets. We down-sampled utterance data (reducing derhotic tokens) at the level of the participant, so long as the participant had more than 200 tokens in the analysis. The down-sample ratio in test and validation was based on the ratio of derhotic:rhotic tokens expected in clinical use with stimulable participants, to maximize replicability. This ratio was estimated to be 2:1 from 229,934 previous practice trials with computerized intervention [18, 19]. The training subset was down-sampled (reducing derhotic tokens) to achieve a 1:1 class ratio. The details regarding each dataset are shown in Table 1.

### 3.2.1. Age and Sex Fairness Exploration

We also stratified the test/validation sets by age and sex to explore the impact of main effects/interactions of these variables on performance. Assignment of participants to test and validation (in that order) was randomized from the set of participants meeting the above criteria, stratified by available ages and sexes. All remaining participants, including those with typical speech, went to the training subset which was not balanced with regard to age and sex. The number of participants selected for each set was fixed to approximate a 70:15:15 token-level balance following down-sampling.

Table 1: *Dataset characteristics. n: number, SSD: speech sound disorder, F: female, Age: $\bar{x}$ ($\sigma_x$) [range], in years.*

| Subset | nSSD /Total | nF | Age | n Derhotic Utterances | n Fully Rhotic Utterances |
|---|---|---|---|---|---|
| Train | 192/303 | 142 | 11.4 (3.5) [6-36] | 36,979 | 32,705 |
| Valid. | 22/22 | 7 | 10.5 (1.7) [8-14] | 5,179 | 9,626 |
| Test | 26/26 | 8 | 10.8 (3.4) [7-24] | 4,849 | 9,183 |

## 3.3. Feature Extraction

All PERCEPT-R v2.2.2 utterances are mono channel, 44.1 kHz .wav audio with an utterance-average intensity of 70 dB.

### 3.3.1. Estimation of Rhotic Interval in Utterance

The spectral segment associated with the target rhotic phoneme in each word was identified using the Montreal Forced Aligner [20] wrapper to Kaldi [21]. Rhotic interval timestamps were estimated using pretrained adult American English acoustic models that were adapted to the PERCEPT-R corpus and a researcher-extended ARPABET dictionary. The rhotic-associated segments were extracted based on these timestamps along with a 10ms buffer extending the size of the interval in each direction to offset edge effects. To assess automated performance, no aligned boundaries were hand corrected.

### 3.3.2. Formant Estimation and Normalization

Formants were extracted from these rhotic-associated intervals using the Praat "To Formants: Robust" algorithm using the Parselmouth API for Python. Five formants were estimated using 5 ms Gaussian-like windows with a 5 ms step between the centers of analysis frames, with preemphasis occurring above 50 Hz. Praat-default values for robust [22] refinement of formant frequencies were used. LPC coefficients were calculated using the Burg algorithm [23]; however, the ceiling of the formant search range was customized for each participant [24]. Ceiling estimation differed between the training subset and the validation/test subsets. For the training set, ceilings were optimized at the utterance level using grid search, minimizing the residuals of the modeled estimate for a given formant, by a custom parallelization-wrapped FastTrack [25]. The first search space was relatively unconstrained (4500-7500 Hz; 500 Hz steps). The participant-specific interquartile range of ceilings was obtained and used to re-constrain participant-search space during a second search (250 Hz steps). These utterance-level ceiling values were passed to Praat for formant estimation. A total of 69,340 training set utterances had formant ceilings determined in this way. For the 344 utterances for whom FastTrack failed, participant-specific average ceilings were passed to Praat. In the validation and test subsets, formant ceilings were determined manually for each participant by observing formant tracks generated with different ceiling values as in [26], in order to mimic the end use methodology of the clinical tool; explorations showed the manual method enhanced performance. No extracted formants were hand corrected.

We retained the first three formants (F1, F2, F3) from these estimates, and calculated F3-F2 distance and F3-F2 deltas. For the age-and-sex normalized feature set, features were z-standardized relative to /ɹ/ values from age-and-sex strata representing formant values from typical speakers in a third party dataset [16] as in [26]. A comparison feature set was generated with features z-standardized at the level of individual rhotic intervals. The following methods were used for both formant feature sets: missing values were imputed using the mean of adjacent estimates in the formant-timeseries. Each timeseries was standardized into 10 bins, such that each bin represented the mean, median, standard deviation, variance, skew, and kurtosis for 10% of the timeseries, with no overlap. This resulted in a 3D array for neural network input: 10 time windows x 5 formants/transforms, and 8 aggregated features. The means for each formant/transform-time window for a given utterance were flattened for testing with shallow classifiers.

*3.3.3. MFCC Estimation*

MFCCs were estimated in analogous fashion, using the Praat "To MFCCs" algorithm. Window length, timestep, and z-standardization were identical to those described above. 13 MFCCs were computed with default filter bank parameters. Timeseries standardization was also identical, yielding a 3D array (10 time windows x 13 MFCCs, and 8 aggregated features) that was flattened for testing with shallow classifiers.

### 3.4. Classifier Architectures, Training, and Tuning

The performance of three neural network architectures (convolutional neural network, CNN; gated recurrent neural network, GRNN; and a CNN-GRNN) were compared with a random forest shallow classifier and a stochastic gradient descent classifier using Pytorch and Scikit-Learn. We hypothesized that the CNN-GRNN would best capture the frequency differences between the fully rhotic and derhotic time series. Shallow classifiers were included as a preliminary comparison of feature sets, as well as monitoring neural network performance with regard to data size sufficiency and possible subset data leakage. The F1-score was the performance metric. In addition to being commonly used, F1-score has the added benefit of indexing recall, which is clinically desired when advancing to higher levels of linguistic complexity during practice.

Training for deep networks was constrained to 25 epochs with early stopping after 5 epochs without decreasing validation loss. Classifier hyperparameters for deep and shallow models alike were tuned using 50 trials in Optuna to maximize participant-specific F1-score mean in the validation set. Tuning constraints for neural networks appear in Table 2. Class-decision thresholds were determined for each trained model based on a grid search maximizing validation F1-score in output. Training occurred on 32 CPU cores and 4 GPU cores split between [*redacted*], with job scheduling managed by HTCondor [27].

Table 2: *Hyperparameter tuning for neural networks*

| Parameter | Possible Values |
|---|---|
| Dropout | $.2 \leq x \leq .5$ |
| Learning Rate/Weight Decay | $1e-5 \leq 1e-1$ |
| Neuron Type | ReLU, GELU, Sigmoid, Tanh, Hardswish, ELU, Hardsigmoid, RReLU, LogSoftmax |
| Optimizer | Adam, RMSprop, SGD, ASGD |
| CNN/GRNN Layers | $1 \leq x \leq 4$ |
| Neurons CNN/GRNN | $16 \leq x \leq 1024$ |
| Linear Layers | $1 \leq x \leq 4$ |
| Neurons Linear Layers | $8 \leq x \leq 1024$ |

### 3.5. Out of the Box Testing and Participant Personalization

First, the trained model from each neural network architecture was applied to the test subset using the out-of-the-box hyperparameters and model decision threshold that maximized performance in the validation set. Next, to reflect the customization possible in a clinical setting, the candidate algorithms were personalized for each participant in the test subset. These data were re-partitioned into re-training, re-validation and test sets using 5-fold cross validation. The hyperparameters for training were identical to those maximizing the validation set for each architecture except that the learning rate was fixed at 1e-3, weight decay was fixed at 0, class-decision thresholds were updated for each participant, and batch size was lowered. Gradients were fixed for all but the last two layers of each neural network. 100 participant-specific trees were added to the random forest and support gradient descent classifiers were trained for an additional 10 epochs.

## 4. Results

### 4.1. Formants Outperform MFCCs

Age-and-sex normalized formants were compared to utterance-normalized formants and MFCC feature sets using stochastic gradient descent and a random forest of decision trees. Each was tuned individually; Table 3 shows average performance.

Table 3: *F1-scores ($\bar{x}$, $\sigma_x$). a/s = age-and sex; u=utterance*

| Feature | Validation Performance | Test – Out of Box | Test – Personalized |
|---|---|---|---|
| MFCCs (u-norm) | .66 (.13) | .68 (.16) | .69 (.17) |
| Formants (u-norm) | .63 (.10) | .66 (.09) | .70 (.10) |
| Formants (a/s-norm) | .73 (.11) | .74 (.13) | **.77 (.12)** |

### 4.2. GRNN Maximizes Personalization versus Other NNs

Our primary outcome is the mean participant-specific F1-score (each themselves the mean of 5-fold cross validation within participant) for the prediction of perceptual judgement of "fully rhotic" or "derhotic" /ɹ/, after personalization to yet-unseen test participants. Due to the preliminary results in 4.1, models were only trained on age-and-sex normalized formant features. The GRNN outperformed other architectures after personalization (Figure 1; Table 5). The GRNN architecture maximizing performance in the validation set had one GRNN layer with 160 neurons and four fully connected linear layers with 191 ReLU neurons followed by dropout.

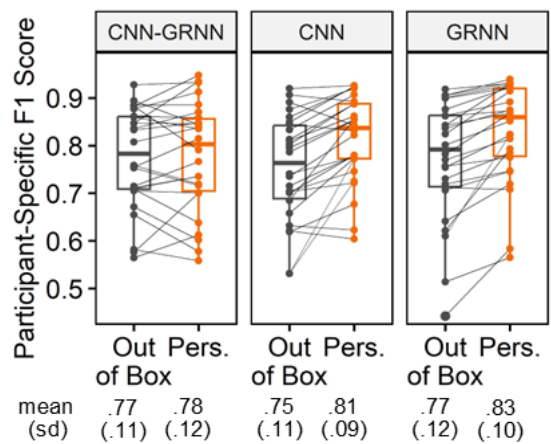

Figure 1: *F1-Scores in the test set using the out of box classifier and following personalization (Pers.)*

### 4.3. Replication: Swapping Validation and Test

Because the stimulability criteria for the test and validation sets favored participants with many tokens, these sets represents ~30% of the token total but only ~15% of the total participants. We also noticed that test consistently outperformed validation. Therefore, to increase external validity, we retrained/retuned a GRNN on the age-and-sex normalized formant feature set after swapping validation and test (Table 4; Table 5). Furthermore, we noticed that in some cases personalization lowered performance versus out of box testing for five IDs. Because we would be able to catch these cases in clinical use, we present a Final condition that represents the best performance per speaker from the Out of Box and Personalized conditions. We present the average of the replications as our final result for the research question at hand. In all experiments, the median value was ≥ than the mean value presented in the tables (med$_{originalfinal}$ = .86, med$_{replicationfinal}$ = .80, med$_{combinedfinal}$ = .83).

Table 4: *Mean and standard deviations of participant specific F1-scores: age-and-sex normalized formants with GRNN; CV = within-participant cross-validation*

| Experiment | Validation Performance | Test Out of Box | Test (CV) Personalized | Test Final |
|---|---|---|---|---|
| Original | .75 (.11) | .77 (.13) | .83 (.08) | .83 (.11) |
| Replication | .80 (.09) | .75 (.11) | .79 (.09) | .80 (.09) |
| **Combined** | .77 (.10) | .76 (.12) | .81 (.10) | **.81 (.10)** |

Table 5: *Participant-weighted confusion matrix for final, combined experiment ($\bar{x}$ = .81 ($\sigma_x$ = .10; med = .83, n = 48), standardized by ground truth.*

| GRNN Prediction | Ground Truth Derhotic | Ground Truth Fully Rhotic |
|---|---|---|
| Derhotic | .70 | .30 |
| Fully Rhotic | .12 | .88 |

### 4.4. Exploration of Age and Sex Model Fairness

The validation and test sets were hand crafted to compare model performance for male and female participants meeting our stimulability criteria at a range of ages. This exploration helps inform whether we should further develop demographic-specific methods for end use in the clinic. We fit a linear mixed model using lme4 in R on the combined test dataset (n = 48 left-out participants) to explore the fixed effects of age, sex, and all interactions on token-level prediction accuracy (Figure 2). Participant-specific intercepts were included as random effects. No fixed effects or interactions were significant at α = .05.

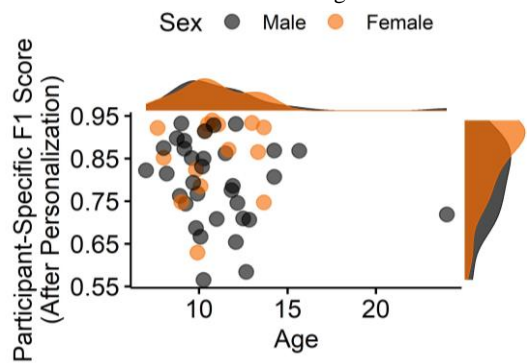

Figure 2: *Post hoc linear mixed modeling of F1-scores show no significant effect of age or sex.*

## 5. Discussion

These results demonstrate the advantages of age-and-sex-normalize formant features for mispronunciation detection of /ɹ/. This work advances the test-participant-specific F1-score for state-of-the-art rhotic classification in the context of stimulable participants with speech sound disorder from $\bar{x}$ = .64 ($\sigma_x$ = .25) [9] $\bar{x}$ = .81 ($\sigma_x$ = .10; med = .83, n = 48).

We found that age-and-sex normalization of formants greatly improved classifier accuracy in shallow classifiers versus utterance-normalized formants and utterance-normalized MFCCs, supporting our hypothesis. Out-of-box testing was comparable between shallow classifiers and neural networks, which we interpret as evidence of replicability for the neural networks. We also interpret the lack of large differences between tested neural network architectures as evidence for the saliency of age-and-sex normalized formants and formant transforms for the detection of rhoticity. Further work can determine if CNN filters can extract these features from the raw audio signal, requiring less hands-on customization for each speaker when creating feature sets.

The lower F1-score during replication might possibly be due to audio or participant differences between validation and test. Acoustician confidence in formant ceiling estimation was informally lower for validation, possibly due to participant-specific audio issues at the time of data collection. Future work can explore these differences as well as compare other test sets

The stratification of our test set allows for statistical evidence that the GRNN predictions were not significantly influenced by sex or age across the range covered by the test set. We interpret this as evidence that the classification performance attained here can extend to individuals who present to the clinic during prospective validation, who are demographically and/or acoustically similar to those in the validation set or test set.

One limitation of the present work is the lack of space to describe the completed SHAP feature importance analysis, which will be reported in longer-format venues. Note that a strength of this study is that we did not exclude tokens with non-unanimous ratings, as has been done in previous investigations of mispronunciation detection [7] and sociophonetic rhoticity [28]. While omitting these tokens would likely increase performance—exploration in our dataset shows $\bar{x}_{F1-score}$ = .88, out of the box—these tokens are clinically encountered and the exclusion of these tokens during model development would have lowered the external validity of this experiment.

## 6. Conclusions

This study presents an age-and-sex normalized formant extraction methodology for the classification of fully rhotic vs derhotic /ɹ/ in the context of speech sound disorder mispronunciation detection. The lab-tested GRNN trained on these features outperformed the lab-tested baseline participant-specific average F1-score from the literature [9] by 17 points ($\bar{x}$ = .81, $\sigma_x$ = .10, med = .83, n = 48).

## 7. Acknowledgements

This research was supported through funding (CUSE II-14-2021; J. Preston, PI) and computational resources (NSF ACI-1341006; NSF ACI-1541396) provided by Syracuse University, and by the National Institute on Deafness and Other Communication Disorders (NIH R01DC017476-S2; T. McAllister, PI).